\documentclass[10pt,twocolumn,british,tightenlines%
,floats,aps,amsmath,amssymb,nofootinbib,superscriptaddress,
prd,showpacs,showkeys,groupedaddress]%
{revtex4-1}
\usepackage{array}
\usepackage{booktabs}
\usepackage{tabu}
\usepackage{dcolumn}
\usepackage{subfigure}
\usepackage{graphicx}% Include figure files
\usepackage{dcolumn}% Align table columns on decimal point
\usepackage{bm}% bold math
\usepackage{slashed}
\usepackage{color}
\usepackage[normalem]{ulem}
\usepackage{babel}
\usepackage{amsmath}
\usepackage{amssymb}
\usepackage{graphicx}
\usepackage{esint}
\usepackage[unicode=true,pdfusetitle,
 bookmarks=true,bookmarksnumbered=false,bookmarksopen=false,
 breaklinks=false,pdfborder={0 0 1},backref=false,colorlinks=false]
 {hyperref}
\usepackage{mathtools}
\makeatletter
\@ifundefined{textcolor}{}
{%
 \definecolor{BLACK}{gray}{0}
 \definecolor{WHITE}{gray}{1}
 \definecolor{RED}{rgb}{1,0,0}
 \definecolor{GREEN}{rgb}{0,1,0}
 \definecolor{BLUE}{rgb}{0,0,1}
 \definecolor{CYAN}{cmyk}{1,0,0,0}
 \definecolor{MAGENTA}{cmyk}{0,1,0,0}
 \definecolor{YELLOW}{cmyk}{0,0,1,0}
}
\usepackage{euscript}\usepackage{epsfig}
\usepackage{amsfonts}\usepackage{fancyhdr}
\usepackage{footmisc}
\makeatother
\begin{document}
\title{Towards Amplituhedron via
       One-Dimensional Theory}

\author{Maysam Yousefian}
        \email{M\_Yousefian@sbu.ac.ir}

\author{Mehrdad Farhoudi }
        \email{m-farhoudi@sbu.ac.ir}
\affiliation{Department of Physics,
             Shahid Beheshti University G.C., Evin, Tehran 19839, Iran}

\begin{abstract}
\noindent Inspired by the closed contour of momentum conservation
in an interaction, we introduce an integrable one-dimensional
theory that underlies some integrable models such as the
Kadomtsev-Petviashvili ({\bf KP})-hierarchy and the amplituhedron.
In this regard, by defining the action and partition function of
the presented theory, while introducing a perturbation, we obtain
its scattering matrix (S-matrix) with Grassmannian structure. This
Grassmannian corresponds to a chord diagram, which specifies the
closed contour of the one-dimensional theory. Then, we extract a
solution of the KP-hierarchy using the S-matrix of theory.
Furthermore, we indicate that the volume of phase-space of the
one-dimensional manifold of the theory is equal to a corresponding
Grassmannian integral. Actually, without any use of supersymmetry,
we obtain a sort of general structure in comparison with the
conventional Grassmannian integral and the resulted amplituhedron
that is closely related to the Yang-Mills scattering amplitudes in
four dimensions. The proposed theory is capable to express both
the tree- and loop-levels amplituhedron (without employing hidden
particles) and scattering amplitude in four dimensions in the
twistor space as particular cases.
\end{abstract}
\date{July 8, 2021}
\maketitle
\section{Introduction}

One of the key points to
understand particle physics is the scattering amplitudes. On the
other hand, the standard method of calculating these scattering
amplitudes is to apply the Feynman diagrams, which have many
complications~\cite{Kampf-2013,Elvang-2015,Kampf-2019}. These
complications have always been a motivation for physicists to
simplify their calculations. A specific example of such efforts is
the Park-Taylor formula, which describes the interaction between
gluons~\cite{Parke-1986,Mangano-1991}.

As another example, in Ref.~\cite{Britto-2005}, it has been shown
that one can, with correct factorization, give a recursive
expression in terms of amplitudes with less external particles, in
which a scattering amplitude is broken down into simpler
amplitudes. Also recently, in Ref.~\cite{Arkani-Hamed-2014}, by
employing the recursive expression used in Ref.~\cite{Britto-2005}
and the twistor theory, a geometric structure has been proposed
that is coined amplituhedron. This structure enables
simplification of the computation of scattering amplitudes in some
quantum field theories such as super-Yang-Mills theory. The
amplituhedron challenges spacetime locality and unitarity as chief
ingredients of the standard field theory and the Feynman diagrams.
Hence, due to spacetime challenges at the scale of quantum
gravity~\cite{Callender-2001} and since locality will have to be
remedied if gravity and quantum mechanics ought to coexist, the
amplituhedron is of much interest. However, the presented
amplituhedron requires supersymmetry and introduces the scattering
amplitude of massless particles, nevertheless no trace of
supersymmetry has been seen up to now, and also each material
particle has mass. On the other hand, it has~not been
specified/discussed why the number of dimensions should be exactly
four in the amplituhedron and, in general, the scattering
amplitude issues.

In this work, we show that interesting integrable models can have
a common foundation and being raised from a single root. For this
purpose, inspired by the fact that geometric representation of the
momentum conservation of participated particles in an interaction
leads to the formation of a closed contour, we introduce a theory
in one dimension that underlies some integrable models in various
dimensions. The important kinds of integrable models that we are
considering here are the KP-hierarchy~\cite{Jimbo-1983},
two-dimensional
models~\cite{Ishibashi-1987,Vafa-1987,Verlinde-1987}, the
amplituhedron~\cite{Arkani-Hamed-2014}. In the proposed theory, as
there is only one dimension, in analogy with the closed contour
resulted from the momentum conservation, there is only one point
that evolves in itself, hence we refer to this theory as a point
theory ({\bf PT}).

In the following, we first define the corresponding action, and
then its equation of state and path integral. Also, we present a
specific symmetry of the PT. Afterward, by defining S-matrix of
the PT, we show the Grassmannian structure of the S-matrix of the
PT, and in turn, we specify its closed contour. Thereupon, we
define a hierarchy transformation by which, we represent some
integrable models based on the PT. Thus, we express a vacuum
solution in the PT as a solution of the KP-hierarchy. Accordingly,
under these considerations, we endeavor to move towards the
amplituhedron without employing supersymmetry. For this purpose,
we indicate the tree and loop levels of the amplituhedron as
special cases of the contour integrations of the volume of
phase-space of the closed contour, wherein the volume itself is as
a Grassmannian integral. Finally, we discuss the physical meaning
and importance of the linked closed contour.

\section{Foundation of Point Theory}

It is convenient to work in the natural units (i.e., $c=1=\hbar$)
with an action on a real $1$-dimensional compact Hausdorff
manifold, say $\mathcal{C}$ as a vacuum, namely
\begin{equation}\label{action}
S=\frac{1}{\kappa}\oint_\mathcal{C} dS(z)\equiv-\frac{1}{\kappa}\oint_\mathcal{C}\psi(z)\varphi(z)dz,
\end{equation}
where $ \psi(z) $ and  $ \varphi(z) $ are two fermionic
fields\footnote{Actually, these two fields are~not independent, as
it will be shown through the variation process.}\
 over the manifold, $\kappa$ is a constant and $z$ is the parameter
of the manifold $\mathcal{C}$.

Each real $1$-dimensional compact Hausdorff manifold (e.g., knots
and links), generally, has its own global topological properties.
For instance, one of these features has been expressed in the
Fary-Milnor theorem~\cite{Sullivan-2008}, which (using a specified
local scalar curvature at any point on the curve) explains whether
or not a $1$-dimensional manifold is a knot. However, global
topological properties of compact manifolds cannot be determined
by local coordinates. For this reason, non-local or
higher-dimensional coordinates, e.g. complex numbers, can be used
instead of real local coordinates to express the global
properties.

Therefore, if the corresponding manifold is a circle, as real
projective line, then its corresponding parameter will at least be
a unit complex number. In the case that the manifold $\mathcal{C}
$ is an unknot and unlinked disconnected compact manifold (e.g.,
several circles), then the parameter $z$ can at least be a complex
number. Also, it is known that a knot is an embedding of a circle
into $3$-dimensional Euclidean space
$\mathbb{R}^3$~\cite{Cromwell-2004} and $\mathbb{R}^3$ is a
subspace of $1$-dimensional biquaternion\footnote{A biquaternion
number $Q$ is a kind of number that is defined as $Q:=\left\lbrace
a_0+\sum_{i=1}^3a_ie_i\left| \right. \lbrace e_i,e_j\rbrace
=-2\,\delta_{ij};a_0,a_i\in\mathbb{C} \right\rbrace $, where
$e_i$'s are quaternion basis. }\
 space. Now, if the
manifold $ \mathcal{C} $ is a knotted and/or a linked one, then
its parameter $z$ will be a member of a special subspace (say,
$\mathbf{M}$, which corresponds to Minkowski space wherein
$\mathbb{R}^3\subset\mathbf{M} $) of biquaternion space
having\footnote{The sign $\star$ on $Q^\star$ is the quaternion
conjugate and is defined as $Q^\star\equiv
a_0-\sum_{i=1}^3a_ie_i$, whereas $Q^*$ is the complex conjugate of
$Q$, i.e. $Q^*=a_0^*+\sum_{i=1}^3a_i^*e_i$.}\
 $Q^*=Q^\star$,
with an additional constraint on the $\mathbf{M}$. An additional
constraint is required because the subspace $\mathbf{M}$
corresponds to $4$-dimensional Minkowski space that has one
dimension more than $\mathbb{R}^3$.

Before we determine such a constraint, first we write each number
of the subspace $\mathbf{M}$ (say, $Q$) in the unitary matrix
representation in terms of the Euler parameters, namely
\begin{equation}\label{QuatMatr}
Q=\begin{pmatrix}
    a_0+ia_3  &  ia_1+a_2 \\[0.3em]
    ia_1-a_2 &  a_0-ia_3
\end{pmatrix}.
\end{equation}
In this way, due to the twistor theory~\cite{Huggett-1994}, each
point (number) in the subspace $\mathbf{M}$ determines a line in
the null projective twistor space, in which each twistor, say
$\varpi$, has coordinates
$\varpi\equiv((\zeta_1,\zeta_2),(\zeta_3,\zeta_4))\equiv(\lambda,\mu)$.
Also, one can specify a point in the null projective twistor space
by two points or a null line in the $\mathbf{M}$. Second, we
determine $n$ different values of the parameter $z$, as different
numbers (points) $z_i\in\mathbf{M}$ for $i=1,\cdots,n$, which fix
the manifold $\mathcal{C}$ (see, e.g., Fig.~$1$). Now, due to
\begin{equation}\label{CoorTwisBiqua}
\begin{cases}
\lambda_iz_i=\mu_i \\ \lambda_iz_{i+1}=\mu_i
\end{cases}
\qquad\mathrm{and}\qquad
\begin{cases}
\lambda_iz_i=\mu_i \\ \lambda_{i+1}z_{i}=\mu_{i+1},
\end{cases}
\end{equation}
we choose the constraint such a way that
\begin{equation}
||(z_i-z_{i+1})||^2\equiv (z_i-z_{i+1})(z_i-z_{i+1})^\star=0.
\end{equation}
\begin{figure}[htbp]
\includegraphics[width=7cm]{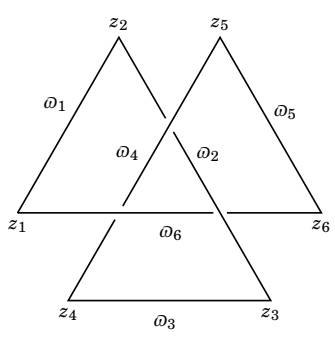}
\caption{\small A $1$-dimensional manifold in $\mathbf{M}$.}
\end{figure}

As mentioned earlier, the 1-dimensional manifold corresponds to
the closed contour resulted from the momentum conservation of
participated particles in an interaction. In this regard, the
closed contour can be considered as a knot, and hence, the
topological properties of the knot can be related to some of the
interaction properties. On the other hand, the closed contour can
be a link. However in this research, we explain the basis of the
PT and its relation with the scattering amplitude and
amplituhedron, and will examine the characteristics of the states
as knot and link in another research. Nevertheless, in the
following, we briefly discuss the physical interpretation of the
states as linked closed contour.

The variation of action \eqref{action} gives two trivial solutions
\begin{equation}\label{eq-of-mo1}
\frac{\delta\partial_zS(z)}{\delta\varphi(z)}=0=\frac{\delta\partial_zS(z)}
{\delta\psi(z)}\ \Rightarrow\ \psi(z)=0=\varphi(z),
\end{equation}
as the Euler-Lagrange equations. However, there is also a
non-trivial solution that, without loss of generality about the
constant of integration, can set to be
\begin{equation}\label{eq-of-mo}
\delta\partial_zS(z)=0\Longrightarrow\psi(z)\varphi(z)=1.
\end{equation}
This solution is the continuity equation in one dimension wherein
$\psi(z)\varphi(z)$ is the conserved current within it.
Considering the non-trivial case, we choose an arbitrary solution
\begin{equation}\label{class-bozoni}
\psi(z)\equiv e^{S(z)}\xRightarrow{\eqref{eq-of-mo}} \varphi(z)= e^{-S(z)},
\end{equation}
which, after quantization, it will be clear that is a good choice.
Also, as there is no derivative of the field in the Lagrangian,
the field momentum is zero. Thus, the energy density is
\begin{equation}\label{ener-den}
h(z)=\psi(z)\varphi(z),
\end{equation}
which, due to Eq.~\eqref{eq-of-mo}, is a
constant\rlap.\footnote{As the fields $\psi$ and $\varphi$ have no
momentum, and hence no dynamics, the one dimensional manifold of
the PT is classically without length.}\
 Also, for later on uses, we expand $ \psi(z) $ and $
\varphi(z) $ as\footnote{In the continuation of this section and a
few later sections, we employ the same formulation mentioned in
Refs.~\cite{Miwa-2000,Jimbo-1983}. However, the lack of the
standard structure of a field theory (such as Lagrangian and field
equations) is the main difference. Besides,
Refs.~\cite{Miwa-2000,Jimbo-1983} have been formulated only in
real one dimension.}
\begin{eqnarray}\label{Lau-exp-2}
&\psi(z)=\sum\limits_{p\in \mathbb{Z}+1/2} \psi_p z^{-(p+1/2)},\cr
&\varphi(z)=\sum\limits_{p\in \mathbb{Z}+1/2} \varphi_p
z^{-(p+1/2)},
\end{eqnarray}
where $\psi_p$ and $\varphi_p$ are coefficients of the expansion.

At this stage, similar to complex numbers, we extend the Cauchy
integral theorem to any biquaternion number $z\in\mathbf{M}$. In
this regard, for each real number $l\neq 1$, we have
\begin{equation}\label{knot-integral1}
\oint_{\mathcal{C}} z^{-l}dz=-(l-1)^{-1}\oint_{\mathcal{C}}
d(z^{-l+1})\mathrel{\mathop=^{l\neq 1}}0.
\end{equation}
When $l=1$, it is known that one can write the point $z$ with
arbitrary parameters $x,y\in \mathbb{R}$ as
\begin{equation}
z= x+y\mathbf{n},
\end{equation}
where $\mathbf{n}=\sum_{i=1}^3a_ie_i$ with $\mathbf{n}^2=1 $. In
this case, using the Stokes-Cartan theorem~\cite{cartan--1945},
the Wick rotation method~\cite{Wick,Peskin}, and transformations
$\mathbf{i}=-h \mathbf{n}$ and $\bar{y}=hy $, wherein $h^2=-1$,
then $z=x+\bar{y}\mathbf{i}$. Hence, we get
\begin{equation}\label{knot-integral2}
    \begin{split}
&\oint_{\mathcal{C}} z^{-1}dz=\int_{\mathcal{S}}dzdz^\star\partial_{z^\star}\partial_z \ln{z} \\
 &=\int_{\mathcal{S}-\mathcal{S}_0}\!\!\!\!\!\!\! dzdz^\star\partial_{z^\star}\partial_z \ln{z}
 +\oint_{\mathcal{C}_0} z^{-1}dz,
    \end{split}
\end{equation}
where $\mathcal{S}$ is the Seifert surface that corresponds to the
closed curve $\mathcal{C}$ containing the point $z=0$,
$\mathcal{S}_0$ is the infinitesimal closed neighborhood of $z=0$
in the Seifert surface, and $\mathcal{C}_0$ is the boundary of
$\mathcal{S}_0$. As $\partial_{z^\star}\partial_z \ln{z}$ on the
surface $\mathcal{S}-\mathcal{S}_0$ is zero, its integration over
the surface $\mathcal{S}-\mathcal{S}_0$ vanishes. Therefore, the
only remaining term of relation \eqref{knot-integral2}, as a
contour integration over a general closed contour (e.g., knotted
or linked closed contour), is the last term. That is, we
have~\cite{Gentili-2007}
\begin{equation}\label{knot-integral3}
        \oint_{\mathcal{C}} z^{-1}dz=\oint_{\mathcal{C}_0} z^{-1}dz= 2\pi
        \mathbf{i}_0,
\end{equation}
where, by $\mathbf{i}_0$, we mean a specific choice of
$\mathbf{i}$ that, without loss of generality, we choose it to be
$\mathbf{i}_0=e_1$.

Accordingly, by employing \eqref{action}, \eqref{Lau-exp-2} and
\eqref{knot-integral3}, the corresponding path integral of the PT
can be written as
\begin{equation}\label{path-integ}
\mathcal{Z}=\int D\psi D\varphi\;
e^{\sum\limits_p\psi_p\varphi_{p}},
\end{equation}
where we have set the constant $\kappa$ in action \eqref{action}
equal to $2\pi  e_1$, $ D\psi
=\prod\limits_{p\in \mathbb{Z}+1/2}d\psi_p $ and $ D\varphi
=\prod\limits_{p\in \mathbb{Z}+1/2}d\varphi_p $. Then, using the
path integral~\eqref{path-integ} and the Berezin
integral~\cite{Berezin-1966}, we can calculate the many-point
correlation function. Alternatively, to calculate the many-point
correlation function, we can use the canonical quantization of $
\psi_p $ and $ \varphi_p $. In this way, we have the Clifford
algebra with fermionic operators $ \hat{\psi}_p $ and $
\hat{\varphi}_p $ that satisfy the anti-commutators
\begin{equation}\label{Cliff-Alj}
\{\hat{\psi}_p,\hat{\psi}_q\}=0,\quad
\{\hat{\varphi}_p,\hat{\varphi}_q\}=0 \quad{\rm and} \quad
\{\hat{\psi}_p,\hat{\varphi}_q\}=\delta_{p,-q},
\end{equation}
where $p$ and $q\in \mathbb{Z}+1/2$. To give a representation of
this algebra, we define the vacuum state as
\begin{eqnarray}\label{vacu-sta}
\hat{\psi}_p|0_\mathcal{C}>=0\;\;(p>0),\;\;\hat{\varphi}_p|0_\mathcal{C}>=0\;\;(p>
0), \cr   <0_\mathcal{C}|\hat{\psi}_p=0\;\;(p<
0),\;\;<0_\mathcal{C}|\hat{\varphi}_p=0\;\;(p < 0).
\end{eqnarray}
Using quantum forms of expansions \eqref{Lau-exp-2}, the quantum
form of the Lagrangian is
\begin{equation}\label{Lau-exp-3}
\partial_z\hat{S}(z)=-\sum\limits_{l\in \mathbb{Z}}\hat{S}_{l-1}z^{-l},
\end{equation}
where
\begin{equation}\label{fermi}
\hat{S}_{l-1}=\sum\limits_{p\in
\mathbb{Z}+1/2}:\hat{\psi}_p\hat{\varphi}_{l-p-1}:,
\end{equation}
and it is compatible~\cite{Miwa-2000} with the quantum forms of
solutions~\eqref{class-bozoni}, namely
\begin{equation}\label{bosoni}
\hat{\psi}(z)=:e^{\hat{S}(z)}: \quad {\rm and}\quad\
\hat{\varphi}(z)=:e^{-\hat{S}(z)}:.
\end{equation}
In the above relations, $ :\hat{a}\hat{b}:$ is the normal ordered
form of $\hat{a}\hat{b}$.
 \vskip0.5cm

\textbf{Symmetry of Point Theory}:$\,{} $ According to solutions
\eqref{bosoni}, the field $\hat{\psi}(z)$ is equal to its
partition function. Thus, due to the quantization process, it
means that the theory will~not change after quantization. This
symmetry is a special feature of the PT that results the
uniformity of physics in large and small
scales\rlap.\footnote{Also, in Ref~\cite{you-far}, we have shown
that, on macro scales, elastic waves in asymmetric environments
behave analogous to QED.}\
 For instance, the result of such a symmetry makes the physics of a
statistical set of leptons and quarks being similar to physics of
each one of those\rlap.\footnote{In Ref.~\cite{YousefFarh5}, we
have proposed a way toward a quantum gravity based on a `third
kind' of quantization approach, in which that similarity has been
explained.}

\section{S-Matrix and Grassmannian Structure}

To express the S-matrix, as an evolution operator for initial
state $|i_\mathcal{C}>$ and final state $|f_\mathcal{C}>$ while
using the free Hamiltonian and relations \eqref{action},
\eqref{knot-integral1}, \eqref{knot-integral3} and
\eqref{Lau-exp-3}, we obtain
\begin{equation}\label{s-mat}
    \mathbb{S}_{if}\!\!= <\!i_\mathcal{C}|\!:\!e^{-\frac{1}{\kappa}\oint_\mathcal{C}
    \hat{\psi}(z)\hat{\varphi}(z)dz}\!:\!|f_\mathcal{C}\!>=<\!i_\mathcal{C}|\!:\!e^{\hat{S}_0}\!:\!|f_\mathcal{C}\!>.
\end{equation}
In other words, using relations \eqref{Cliff-Alj} and
\eqref{fermi}, one equivalently gets
\begin{equation}\label{comu-re}
    :\!e^{\frac{1}{\kappa}\oint_\mathcal{C}\hat{\psi}(z)\check{\varphi}(z)dz}\!:\hat{\psi}(z'):
    \!e^{-\frac{1}{\kappa}\oint_\mathcal{C}\hat{\psi}(z)\hat{\varphi}(z)dz}\!:=e\,\hat{\psi}(z'),
\end{equation}
where $e$ is the Euler number. That is, we have a trivial case
wherein the initial and final states are the same. However, adding
a perturbation to the free Hamilton results a general nontrivial
expression for the S-matrix in the PT, wherein the initial state
evolves. In this respect, we define a perturbed Hamiltonian as the
interaction of superposition of $n$ specific points of
$\mathcal{C}$ with the other points, namely
\begin{equation}\label{pertu-Hamil}
  \begin{split}
   \hat{\psi}(z)&\hat{\varphi}(z)\rightarrow\hat{\psi}(z)\hat{\varphi}(z)+\\
   & \iota\, C(z)\hat{\psi}(z)\big[\hat{\varphi}(z_1')+\cdots+\hat{\varphi}(z_n')\big],
  \end{split}
\end{equation}
where $\iota$ is a constant and $C(z)$ for $a=1,\cdots,n$ is the
probability function of interaction of $\hat{\psi}(z)$ with one or
more points of $\mathcal{C}$. Accordingly, we express the S-matrix
operator in terms of the Dyson series as
\begin{eqnarray}\label{Dyson}
  \!\hat{\mathbb{S}}&=&\!\!\sum\limits_{k=1}^{n}\!\frac{(\frac{\iota}{\kappa})^{k}}{k!}
  :\!\prod\limits^{k}_{\alpha=1}\oint_\mathcal{C}\! \left[\sum_{a=1}^{n}C_\alpha(z_\alpha)
  \hat{\psi}(z_\alpha)\hat{\varphi}(z_a')\right]\!:dz_\alpha\cr
  &\equiv &\!\!\sum\limits_{k=1}^{n}\hat{\mathbb{S}}_{(k)},
\end{eqnarray}
wherein the constant $\iota$ is determined in a way that
\begin{equation}\label{iota}
    <0_\mathcal{C}|\hat{\mathbb{S}}|0_\mathcal{C}\!>=1,
\end{equation}
and given that the product of two identical Fermion fields is
zero, then $\hat{\mathbb{S}}_{(k)}$ for $k>n$ vanishes.

Each $\hat{\mathbb{S}}_{(k)}$ is equivalent to the simultaneous
interaction of $k$ ordered points of $n$ specific points
$(z_1',\cdots,z_n')$ within $\mathcal{C}$. Also, each
$\hat{\mathbb{S}}_{(k)}$ has a Grassmannian structure with a
Grassmannian $G(k,n)$, say $C$, as
\begin{equation}\label{Grass--struc}
    C\equiv
    \begin{pmatrix}
        C_{1}(z_1') & \cdots & C_{1}(z_n') \\
        \vdots & \ddots & \vdots \\
        C_{k}(z_1')& \cdots & C_{k}(z_n')
    \end{pmatrix},
\end{equation}
in a way that
$<\!0_\mathcal{C}|\hat{\mathbb{S}}_{(k)}|0_\mathcal{C}\!>$ being
equal to the sum of minors of the Grassmannian $C$.

On the other hand, each minor of the Grassmannian $C$ is the
probability of simultaneous interaction of $k$ specific ordered
points of $(z_1',\cdots,z_n')$ within $\mathcal{C}$. Also, due to
the conditional probability in the probability theory, the
probability that each point $z_a'$ of $(z_1',\cdots,z_n')$
interacts with the manifold in the $\alpha$'th order is
\begin{equation}\label{cond-probab}
    \bar{c}_{\alpha a} \equiv\dfrac{\det  \begin{pmatrix}
            c_{1a} & \cdots & c_{1,a+\alpha-1} \\
            \vdots  & \ddots & \vdots  \\
            c_{\alpha a} & \cdots & c_{\alpha,a+\alpha-1}
    \end{pmatrix}}{\det  \begin{pmatrix}
            c_{1a} & \cdots & c_{1,a+\alpha-2} \\
            \vdots  & \ddots & \vdots  \\
            c_{\alpha-1, a} & \cdots & c_{\alpha-1,a+\alpha-2}
        \end{pmatrix}},
\end{equation}
where these two determinants are the probability of simultaneous
interaction of $(\alpha-1)$ ordered points and the same
$(\alpha-1)$ ordered points plus one extra point of $n$ specific
points $(z_1',\cdots,z_n')$ within $\mathcal{C}$, and each
$c_{\alpha a}\equiv C_\alpha(z_a')$ is the element of the matrix
$C$.

Thus, due to the number of minors of the Grassmannian $C$ (that is
equal to the combination $\textbf{C}^n_k$) and given that the sum
of the probabilities of all possible states of the simultaneous
interaction of $k$ ordered points of $n$ specific points
$(z_1',\cdots,z_n')$ within $\mathcal{C}$ is equal to one, by
employing the Wick theorem~\cite{Miwa-2000}
\begin{eqnarray}\label{Wick's}
    &<0_\mathcal{C}|\hat{\psi}
    (z_1)\cdots\hat{\psi}(z_n)\hat{\varphi}(z_1')\cdots\hat{\varphi}(z_n')|0_\mathcal{C}>=
    \cr &\det\left(
    <0_\mathcal{C}|\hat{\psi}(z_i)\hat{\varphi}(z_j')|0_\mathcal{C}>\right),
\end{eqnarray}
and relations \eqref{Lau-exp-2} and \eqref{knot-integral3} that
result
\begin{equation}\label{value}
    <\!0_\mathcal{C}|\frac{1}{\kappa}\oint_\mathcal{C} \hat{\psi}(z)\hat{\varphi}(z')dz|0_\mathcal{C}\!>=1,
\end{equation}
we obtain
\begin{equation}\label{probabi}
    \frac{k!}{\iota^k}<\!0_\mathcal{C}|\hat{\mathbb{S}}_{(k)}|0_\mathcal{C}\!>=
    \sum\limits_{h=1}^{\textbf{C}^n_k}M_h=1.
\end{equation}
Furthermore, the minors $M_h$ of the Grassmannian $C$, as
probability of simultaneous interaction of $k$ specific points of
$(z_1',\cdots,z_n')$ within $\mathcal{C}$, are, in general,
positive numbers. Therefore, the Grassmannian $C$ is a positive
Grassmannian. In addition, in Ref.~\cite{Kodama-2017}, it has been
shown that each non-negative Grassmannian corresponds to a chord
diagram. In this way, as a chord diagram is a closed contour, each
non-negative Grassmannian, and in turn each
$\hat{\mathbb{S}}_{(k)}$, specifies a closed contour
$\mathcal{C}$. Moreover, since a Grassmannian $G(k,n)$ is a space
that parameterizes all $k$-dimensional linear subspaces of the
$n$-dimensional vector space, if we mod out it by the general
linear group $GL(k)$, there will be no~change in it. Then,
dividing relation \eqref{probabi} by one of $M_h$ or a constant,
makes no~change in $\mathcal{C}$. Eventually, it is obvious that
the S-matrix in the PT has a Grassmannian structure.

\section{Hierarchy Transformation and KP-Hierarchy}

To investigate the integrability and solitonic properties of the
PT, we inquire relation between the PT and the KP-hierarchy. The
KP-hierarchy is based on mathematical foundation in terms of a
Grassmannian variety~\cite{Jimbo-1983,Miwa-2000,Kodama-2017}. It
is supplemented with a number of classical and non-classical
reductions that contain many integrable equations and solitonic
models~\cite{Jimbo-1983}. Indeed, the KP-hierarchy is a hierarchy
of nonlinear partial differential equations in a function, say
$\tau(x)$, of infinite number of variables $x_l$ for
$l=1,\cdots,\infty$ as hierarchy coordinates. As the KP-hierarchy
is an integrable system, the plausible (physical) meaning of the
hierarchy coordinates can be related to the infinite numbers of
possible conserved quantities that any KP-hierarchy soliton
solution can contain~\cite{Kodama--2011,Kodama--2014}. Such a
$\tau(x)$ is a function that should satisfy the bilinear
identity~\cite{Jimbo-1983}
\begin{equation}\label{tau}
\oint_\mathcal{C}
e^{\xi(k,x-x')}\tau\big(x-\epsilon(k)\big)\tau\big(x'+\epsilon(k)\big)dk\equiv
0,
\end{equation}
where $\xi(k,x)\equiv\sum_{l=0}^\infty x_lk^l$, $k\in\mathbf{M}$
is an arbitrary point in the manifold ${\mathcal C}$,
$\epsilon(k)\equiv \big(1,1/k,1/(2k^2),\cdots\big)$, and arbitrary
variables $x\equiv (x_0,x_1,x_2,\cdots)$ and $x'\equiv
(x'_0,x'_1,x'_2,\cdots)$ with $x_l,x'_l\in\mathbf{M}$.

To show the relation between the PT and the KP-hierarchy, we
employ a transformation, which we will refer to it as a hierarchy
transformation ({\bf HT}), in the form
\begin{equation}\label{Hier-Transf}
\text{HT:}
\begin{cases}
\hat{\psi}(z)\rightarrow e^{H(x)}\hat{\psi}(z)e^{-H(x)}\cr
\hat{\varphi}(z)\rightarrow e^{H(x)}\hat{\varphi}(z)e^{-H(x)},
\end{cases}
\end{equation}
where
\begin{equation}\label{difeomorph}
    H(x)\equiv \oint_\mathcal{C} \xi(z',x)dS(z')/(2\pi
    e_1).
\end{equation}
Given that the function $\xi(k,x)$, for constant parameters
$x\equiv (x_0,x_1,x_2,\cdots)$, is regular with respect to $k$,
relation \eqref{difeomorph} means that the closed contour
$\mathcal{C}$ in relation \eqref{action}, without changing the
topological properties\rlap,\footnote{Via a smooth variation of
$x\equiv (x_0,x_1,x_2,\cdots)$, the function $\xi(k,x)$ can tend
to one. In this way, $\mathcal{C}'$ smoothly tends to
$\mathcal{C}$.}\
 is subjected to a diffeomorphism as
\begin{equation}\label{difeo}
\begin{split}
        H(x)&=- \oint_\mathcal{C} \xi(z',x)dz'\psi(z')\varphi(z')/(2\pi
    e_1)\\ & =-\oint_{\mathcal{C}'} dz''\psi'(z'')\varphi'(z'')/(2\pi
    e_1),
\end{split}
\end{equation}
where $dz''\equiv\xi(z',x)dz'$,
$\psi'(z'')\varphi'(z'')=\psi(z')\varphi(z')$ and $\mathcal{C}'$
is a diffeomorphism of $\mathcal{C}$. Hence, after some
calculations~\cite{Miwa-2000}, transformation~\eqref{Hier-Transf}
reads
\begin{equation}\label{Hier-Transform}
\begin{cases}
\hat{\psi}(z)\rightarrow
e^{-\xi(z,x)}\hat{\psi}(z)\equiv\hat{\psi}(z,x)\cr
\hat{\varphi}(z)\rightarrow
e^{\xi(z,x)}\hat{\varphi}(z)\equiv\hat{\varphi}(z,x).
\end{cases}
\end{equation}

Now, we define $\hat{\mathbb{S}}_{(k,x)}$ as
\begin{equation}\label{skx}
 \hat{\mathbb{S}}_{(k,x)}\!\!\equiv \!\frac{(\frac{\iota}{\kappa})^{k}}{k!}\!:\!\!\prod\limits^{k}_{\alpha=1}
 \!\oint_\mathcal{C}\!\! \left[\sum_{a=1}^{n}\!\!C_\alpha(z_\alpha)\hat{\psi}(z_\alpha,x)
 \hat{\varphi}(z_a')\right]\!\!:\!dz_\alpha.
\end{equation}
Since, the function $\tau(x)$, as a solution of the KP-hierarchy,
is define as
\begin{equation}\label{tua}
    \tau(x)=\frac{k!}{\iota^k}<\!0_\mathcal{C}|\hat{\mathbb{S}}_{(k,x)}|0_\mathcal{C}\!>=\mathrm{Wr}(f_1,\cdots,f_k),
\end{equation}
where $\mathrm{Wr}(f_1,\cdots,f_k)$ is the Wronskian of functions
$f_\alpha$ for $\alpha=1,\cdots,k $ as each $f_\alpha$ is
\begin{equation}\label{f-func}
    f_\alpha\equiv\left(c_{\alpha 1} e^{-\xi(z_1',x)},\cdots,c_{\alpha n}e^{-\xi(z_n',x)}\right).
\end{equation}

At this stage, as in
Refs.~\cite{Miwa-2000,Chakravarty-2014,Kodama-2017}, it has been
shown that the $\mathrm{Wr}(f_1,\cdots,f_k)$ is a solution of the
KP-hierarchy\rlap,\footnote{Incidentally, in
Refs.~\cite{Ishibashi-1987,Vafa-1987,Verlinde-1987}, it has been
shown that the $n$-point correlation function of fields of a kind
of string theory can be obtained via the KP-hierarchy solutions.}\
 thus we conclude that a vacuum solution in the PT is equivalent to a solution of the
KP-hierarchy.

\section{Phase-Space of $\mathcal{C}$}

Consider a quantum system with a Hamiltonian operator
$\hat{\mathcal{H}}$ and its basis states. The time evolution of
the probability of a state of the system, say $p_i$, in this
Hamiltonian basis is
\begin{equation}\label{evolution}
    \partial_tp_i=i\hat{\mathcal{H}}p_i=i\mathcal{H}_ip_i
\end{equation}
that leads to
\begin{equation}\label{pha-sp}
    d\ln{p_i}=i\mathcal{H}_idt.
\end{equation}
For a one-dimensional quantum system, relation \eqref{pha-sp} is
proportional to the volume element of phase-space of the state
$p_i$. In this way, the volume element of phase-space of total
system is proportional to
\begin{equation}\label{tot-pha-sp}
    \prod\limits_{i}d\ln{p_i}.
\end{equation}

Inspired by this explanation, due to definition
\eqref{cond-probab} and as the Grassmannian $C$ specifies the
closed contour $\mathcal{C}$, we (plausibly) assume that the
volume element of phase-space of the closed contour $\mathcal{C}$
is\footnote{Given that each Grassmannian specifies a chord diagram
and each chord diagram determines a one-dimensional manifold,
hence the number of dimensions of the phase-space of the
one-dimensional manifold is equal to the number of degrees of the
Grassmannian freedom.}
\begin{equation}\label{tot-pha-sp-c}
    \prod\limits^{k, n}_{\alpha, a=1}d\ln{\bar{c}_{\alpha a}}.
\end{equation}
Then, using definition \eqref{cond-probab} and
\begin{equation}\label{min}
    M_b \equiv \det  \begin{pmatrix}
        c_{1,b} & \cdots & c_{1,b+k-1} \\
        \vdots  & \ddots & \vdots  \\
        c_{k b} & \cdots & c_{k,b+k-1}
    \end{pmatrix},
\end{equation}
expression \eqref{tot-pha-sp-c} reads as
\begin{equation}\label{integ-matrix}
\dfrac{d^{k\times
            n}c_{\alpha a}}{\prod\limits_{b=1}^n
        M_b}.
\end{equation}

However, there are two points that must be considered in order to
define the volume element of the phase-space of $\mathcal{C}$.
First, as the Grassmannian $C$ is invariant under general linear
group $GL(k)$, the volume element of the phase-space of it must be
mod out by $GL(k)$ `gauge' redundancy. The second point is about
the relation between the Grassmannian $C$ and the coordinates
$z_a$. Given relation \eqref{CoorTwisBiqua}, twistors, say
$\varpi_a=(\zeta_{a1},\cdots,\zeta_{a4})$, can correspond to the
lines between two points $z_a$ and $z_{a-1}$ in analogy with the
momentum of particles in the closed contour of momentum
conservation. Accordingly, if we define\footnote{We use the
Einstein summation rule, unless it is specified.}
\begin{equation}\label{y}
c_{\alpha a}\varpi_a\equiv y_\alpha,
\end{equation}
two different cases will be possible. One case is that the inner
product of the $4$ $n$-component vector ${\boldsymbol\varpi}$ to
each of the $ k$ $n$-component row-vectors of the $C$ matrix being
zero (i.e., all of $y_\alpha=0$). The other case is that some of
$y_\alpha$'s being zero.

In the above first case, there would be a $(4+k)$-plane. Now, we
employ some auxiliary parameters, say a matrix $\mathtt{z}$ made
of elements $\mathtt{z}_{a\alpha}$, as
\begin{equation}\label{z}
C\cdot\mathtt{z}\equiv\mathtt{y}\qquad\ \text{with
constraint:}\quad \mathtt{y}=\mathbb{I}_{k\times k}.
\end{equation}
Then, due to the property of the matrix $C$, this matrix
$\mathtt{z}$ contains $k$ independent column-vectors, which
(within the first case) are also independent from $4$
$n$-component vector $\varpi_a$. Accordingly, we can choose a
matrix, say $Z$, consists of $n$ row-vectors $Z_a\equiv
(\zeta_{a1},\cdots,\zeta_{a4},\mathtt{z}_{a1},...,\mathtt{z}_{ak})$,
which contains $(4+k)$ independent $n$-component column-vectors as
a basis for the above $(4+k)$-plane. Such a basis and the above
constraint eliminate the gauge redundancy $ GL(k)$. In this case,
we can assume a polygon in a $(3+k)$-dimensional projective space
whose $n$ vertices are $Z_a$'s.

At this stage, we satisfy the elimination of the gauge redundancy
$GL(k)$ and also the application of those two mentioned cases
\eqref{y} by imposing the Dirac delta functions to the volume
element of the phase-space as
\begin{equation}\label{vol-hie}
\dfrac{d^{k\times
            n}c_{\alpha a}}{\prod\limits_{b=1}^n
        M_b}\delta^{k\times
        k}\mathtt{(y-\mathbb{I}_{k\times k})}\delta^{k'\times4}(y-0_{k'\times 4}).
\end{equation}
Indeed, the Dirac delta functions $\delta^{k\times
k}\mathtt{(y-\mathbb{I}_{k\times k})} $ eliminate the gauge
redundancy $GL(k)$ and the Dirac delta functions
$\delta^{k'\times4}(y-0_{k'\times 4})$ impose those two mentioned
cases of \eqref{y}. Therefore, expression \eqref{vol-hie}
indicates that the volume of the phase-space is equal to a
corresponding Grassmannian integral\rlap.\footnote{For definition
of the Grassmannian integral, see, e.g.,
Refs.~\cite{Elvang-2015,Arkani-Hamed-2016}.}\
 In addition, due to
definition $M_b$'s in \eqref{min}, the volume element, and in turn
the corresponding Grassmannian integral, is obviously
cyclic-invariant.

\section{Four-Dimensional Tree-Amplituhedron}

In this section, we endeavor to show that the volume of the
phase-space of $\mathcal{C}$ is equivalent with the amplituhedron
in a four-dimensional spacetime. For this propose, first we
indicate that the tree-amplituhedron in a four-dimensional
spacetime can be deduced from the first case of definition
\eqref{y}. Actually, in this case, the integration of expression
\eqref{vol-hie} becomes
\begin{equation}\label{4-vol-hie}
\int\dfrac{d^{k\times n}c_{\alpha
a}}{\prod\limits_{b=1}^nM_b}\delta^{k\times(4+k)}(\mathtt{Y}-\mathtt{Y}_0)\equiv
\mathcal{A}_{n,k},
\end{equation}
where $\mathtt{Y}$ is a $[k\times (4+k)]$-matrix generally defined
as $\mathtt{Y}\equiv C\cdot Z$ with constraint $\mathtt{Y}_0
=(0_{k\times 4},\mathbb{I}_{k\times k})$. Furthermore, consider
each cell decomposition (say $T$) of the polygon with the $n$
vertices $Z_a$'s, which has a few cells (say $\Gamma$'s), wherein
any cell of it, as a particular example with
Grassmannian\footnote{In fact for these particular examples, the
    components of the corresponding Grassmannian are determined by the
    Dirac delta functions in the Grassmannian integral. And, the rest
    of components of the $C$ matrix in the Grassmannian integral are
    removed by contour integration. In other examples of cells/contour
    integration, the corresponding Grassmannian integral leads to
    other solutions that can principally be obtained in similar
    process of these particular examples. Of course, to avoid
    complicating the issue in this work, we will present some other
    examples in further studies.\label{ffoott}}
\begin{equation}\label{matrix4}
    \begin{pmatrix}
        c_{1,a_{(1)}} & \cdots & c_{1,a_{(4+k)}} \\
        \vdots & \ddots & \vdots \\
        c_{k,a_{(1)}}& \cdots & c_{k,a_{(4+k)}}
    \end{pmatrix},
\end{equation}
is specified as
\begin{equation}\label{CellDecom4}
\mathtt{Y}_\alpha\equiv\sum\limits_{i=0}^{k+3}c_{\alpha,
a_{(1+i)}}Z_{a_{(1+i)}}.
\end{equation}
In here, $\mathtt{Y}_\alpha $ is $\alpha$th row-vector of the
defined matrix $\mathtt{Y}$ but with $(4+k)$ components, wherein
$k\times(n-k-4)$ components of the matrix $C$ have been set to be
zero. Hence, using \eqref{CellDecom4}, each of the corresponding
minors \eqref{min} and the related multiplication of differential
elements for any cell are
\begin{equation}\label{4-1cell-de}
M_{a_{(i)}}=\frac{<\mathsf{Y},Z_{a_{(1+i)}},\cdots,Z_{a_{(4+i)}}>}{<Z_{a_{(1)}},Z_{a_{(2)}},\cdots,
Z_{a_{(4+k)}}>}
\end{equation}
and
\begin{equation}\label{4-diff-1cell-de}
\prod\limits^{4+k}_{i=1} dc_{\alpha
a_{(i)}}=\frac{d^{4+k}\mathtt{Y}_\alpha}{<Z_{a_{(1)}},\cdots,
Z_{a_{(4+k)}}>}.
\end{equation}

Now, by inserting relations \eqref{4-1cell-de} and
\eqref{4-diff-1cell-de} into expression \eqref{4-vol-hie}, the
volume of the phase-space becomes
\begin{eqnarray}\label{4-hier-pha-spa}
&\!\!\sum\limits_{\Gamma\subset T}\!\int \frac{
d^{k\times(4+k)}\mathtt{Y}_\alpha<Z_{a_{(1)}},\cdots,
Z_{a_{(4+k)}}\!>^4\delta^{k\times(4+k)}(\mathtt{Y}\!-\!\mathtt{Y}_0)}
{\!\!\!<\mathsf{Y}\!,Z_{a_{(1)}}\!,Z_{a_{(2)}}\!,Z_{a_{(3)}}
\!,Z_{a_{(4)}}\!>\cdots<\mathsf{Y}\!,Z_{a_{(4+k)}}\!,Z_{a_{(1)}}\!,Z_{a_{(2)}}\!,Z_{a_{(3)}}\!>}=\cr
&\!\!\sum\limits_{\Gamma\subset T} \frac{ <Z_{a_{(1)}},\cdots,
Z_{a_{(4+k)}}>^4}
{\!\!\!<\mathsf{Y}_0,Z_{a_{(1)}}\!,Z_{a_{(2)}}\!,Z_{a_{(3)}}
\!,Z_{a_{(4)}}>\cdots<\mathsf{Y}_0,Z_{a_{(4+k)}}\!,Z_{a_{(1)}}\!,Z_{a_{(2)}}\!,Z_{a_{(3)}}\!>},\cr &
\end{eqnarray}
which is a simple contour integration of Grassmannian
integral~\eqref{4-vol-hie} as a four-dimensional
tree-amplituhedron in the twistor space provided the particular
choice
\begin{equation}\label{bosonization}
    \mathtt{z}_{a\alpha}=\Gamma_a^A\Phi_{A\alpha}
\end{equation}
being selected. In choice \eqref{bosonization}, $\Gamma_a^A$ and
$\Phi_{A\alpha}$ are fermionic variables, and the index $A$ is an
internal fermionic index that represents the number of copies of
supersymmetry. For instance, the simplest case of relation
\eqref{4-hier-pha-spa} is when $k=1$ that, with the particular
choice \eqref{bosonization}, is structurally similar to the
simplest case of amplituhedron in the momentum twistor space
mentioned in
Refs.~\cite{Arkani-Hamed-2014,Arkani-Hamed-2016,Kojima-2020}.
Moreover, relation \eqref{4-hier-pha-spa}, at $k=2$ and $n=6$ with
choice \eqref{bosonization}, is structurally similar to relation
($3.18$) mentioned in Ref.~\cite{Kojima-2020}. Incidentally, as
the Grassmannian integral \eqref{4-vol-hie} is cyclic-invariant,
the resulted relation \eqref{4-hier-pha-spa} from it is
cyclic-invariant as well.

As the scattering amplitude is typically defined in the momentum
twistor space, and the above relation does~not also indicate the
energy-momentum conservation, it is~not suitable to calculate the
scattering amplitude. Thus, by change of variables via the Fourier
transformation of expression \eqref{4-vol-hie}, similar to
Ref.~\cite{Arkani-Hamed-2010}, we transform $\mathcal{A}_{n,k}$
into the momentum twistor space where new results emerge. That is,
we write
\begin{equation}\label{pre-MHV-4-vol-hie}
 \begin{split}
&\int
d^{2n}\mu_a e^{i\tilde{\lambda}_b\wedge\mu_b}\mathcal{A}_{n,k}(\lambda_a,\mu_a,\mathtt{z}_{a\alpha})=\\
& \int\dfrac{d^{k\times n}c_{\alpha a}}{\prod\limits_{c=1}^n
M_c}\delta^{2k}(c_{\alpha b}\lambda_b)\delta^{k\times
k}(\chi)\!\!\!\int
d^{2k}\rho_\alpha\delta^{2n}(\tilde{\lambda}_a-\rho_{\beta}
c_{\beta a}),
 \end{split}
\end{equation}
where we have used the Dirac delta function expression
$\delta^{2k}(c_{\alpha a}\mu_a)=\int d^{2k}\rho_\alpha
\exp{(-i\rho_{\beta} c_{\beta a}\wedge\mu_a)}$ with arbitrary
$\rho_\alpha\equiv(\rho_{\alpha 1},\rho_{\alpha 2})$,
$\chi\equiv\mathtt{y}-\mathbb{I}_{k\times k}$,
$\lambda_a\equiv(\zeta_{a1},\zeta_{a2})$,
$\mu_a\equiv(\zeta_{a3},\zeta_{a4})$ and
$\tilde{\lambda}_a\equiv(\vartheta_{a1},\vartheta_{a2})$ is the
Fourier conjugate of $\mu_a$. Then, in relation
\eqref{pre-MHV-4-vol-hie}, we first integrate over $c_{\mu a}$
for, without loss of generality, $\mu =1,2$ and afterward, over
$\rho_\alpha$. Hence, after some manipulations, it reads
\begin{eqnarray}\label{mini-MHV-4-vol-hie}
&<\!\rho_1,\rho_2\!>^{(k+2-n)}\!\!\delta^4\! \left(
|\tilde{\lambda}_f\!\!><\!\!\lambda_f|\right)\!\int\!\dfrac{d^{\hat{k}\times
n }c_{\hat{\alpha}a}}{\prod\limits_{c=1}^n
M_c}\delta^{2\hat{k}}(c_{\hat{\alpha} b}\lambda_b)\times\cr
&\delta^{\hat{k}\times \hat{k}}(\chi_{\hat{\alpha}
\hat{\beta}})\delta^{2\hat{k}}(c_{\hat{\alpha}
e}\mathtt{z}_{e\mu}),
\end{eqnarray}
where we have used constraint \eqref{z}, the Dirac delta function
property
\begin{equation}\label{delta-chang}
(ad-bc)\delta(ar_1+br_2)\delta(cr_1+dr_2)=\delta(r_1)\delta(r_2),
\end{equation}
for any variables $(r_1,r_2) $ with arbitrary parameters $a, b, c$
and $d$, a part of definition \eqref{y} with constraints
$c_{\hat{\alpha}a}\lambda_a=0$, and $\hat{\alpha},
\hat{\beta}=3,\cdots,k$, $\hat{k}\equiv k-2$, $\chi_{\hat{\alpha}
\hat{\beta}}\equiv c_{\hat{\alpha}
a}\mathtt{z}_{a\hat{\beta}}-\delta_{\hat{\alpha} \hat{\beta}}$,
$\rho_\alpha=\tilde{\lambda}_a \mathtt{z}_{a\alpha}$ and $c_{\mu
a}=\left(<\rho_\nu,\tilde{\lambda}_a>
-c_{\hat{\alpha}a}<\rho_\nu,\rho_{\hat{\alpha}}>\right)\times\epsilon_{\nu\mu}/<
\rho_1,\rho_2>$.

For later on convenient, we introduce a tensor, say ${\bold Q}$,
with components
\begin{equation}\label{change-vari}
Q_{ab}\equiv\frac{\delta_{a-1,a,a+1}^{\rm
cef}<\tilde{\lambda}_c,\tilde{\lambda}_e>
\delta_{fb}}{2<\tilde{\lambda}_{a-1},\tilde{\lambda}_a><\tilde{\lambda}_a,\tilde{\lambda}_{a+1}>},
\end{equation}
where $\delta_{a-1,a,a+1}^{\rm cef} $ is the generalized Kronecker
delta, then the Schouten identity, i.e.
$\epsilon^{abc}<\tilde{\lambda}_a,\tilde{\lambda}_b>\tilde{\lambda}_c\equiv
0$, implies $Q_{ab}\tilde{\lambda}_b=0$. Accordingly, if we
replace $c_{\hat{\alpha}b}$ with
$c_{\hat{\alpha}a}(\delta_{ab}-Q_{aa'}<\tilde{\lambda}_{a'},\tilde{\lambda}_b>)$
in expression \eqref{mini-MHV-4-vol-hie}, it will~not change.
Also, we define $[(k-2)\times n]$-matrix $G$ with elements
$g_{\hat{\alpha}a}$ as $g_{\hat{\alpha}b}\equiv
c_{\hat{\alpha}a}Q_{ab}$, which, for fixed $b$, it implies the
relation between minors \eqref{min} and the minors of matrix $G$
as
\begin{equation}\label{minorminor}
M_b=N_{b+1}\frac{<\tilde{\lambda}_b,\tilde{\lambda}_{b+1}>
\cdots<\tilde{\lambda}_{b+k-2},\tilde{\lambda}_{b+k-1}>}{<\rho_1,\rho_2>},
\end{equation}
where $N_b $'s are independent minors of $G$ as $N_b
=\mathrm{det}[g_{\hat{\alpha},b+\hat{\beta}-1}]$ with the minors
indices $\hat{\alpha}\hat{\beta}$. Then, by employing relation
\eqref{minorminor} and the gauge constraints $c_{\hat{\alpha}
a}\mathtt{z}_{a\mu}=0$, and defining $\tilde{\mu}_{b}$ as
$Q_{ab}\tilde{\mu}_{b}\equiv\lambda_a$ and
$\tilde{\mathtt{z}}_{b\hat{\alpha}}$ as
$Q_{ab}\tilde{\mathtt{z}}_{b\hat{\alpha}}\equiv\mathtt{z}_{a\hat{\alpha}}$
into expression \eqref{mini-MHV-4-vol-hie}, after some
manipulations\rlap,\footnote{For derivations of the relations
mentioned in this section, some techniques, which have been
applied in Ref.~\cite{Arkani-Hamed-2010}, have been used.}\
 it becomes
\begin{equation}\label{MHV-4-vol-hie}
\mathrm{A}^{\mathrm{MHV}}_{n}\int\dfrac{d^{\hat{k}\times n
}g_{\hat{\alpha}a}}{\prod\limits_{c=1}^n
N_c}\delta^{2\hat{k}}(g_{\hat{\alpha}b}\tilde{\lambda}_b)\delta^{2\hat{k}}(g_{\hat{\alpha}e}\tilde{\mu}_e)\delta^{\hat{k}\times
\hat{k}}(\hat{\chi}_{\hat{\alpha} \hat{\beta}}),
\end{equation}
where
\begin{equation}\label{MHV}
\mathrm{A}^{\mathrm{MHV}}_{n}\equiv\frac{<\rho_1,\rho_2>^4\delta^4
\left(|\tilde{\lambda}_b><\lambda_b|\right) }
{\prod\limits^n_{a=1}<\tilde{\lambda}_a,\tilde{\lambda}_{a+1}>}.
\end{equation}

This result is related to the corresponding one mentioned in
Ref.~\cite{Elvang-2015} except that $<\rho_1,\rho_2>^4$ has been
replaced by fermionic delta function. In fact,
definition~\eqref{MHV} can be considered as a sort of general of
the one mentioned in Ref.~\cite{Elvang-2015}, which with the
particular choice~\eqref{bosonization}, is equivalent to the one
mentioned there. Eventually, in analogous with expression
\eqref{4-vol-hie}, expression \eqref{MHV-4-vol-hie} actually reads
\begin{equation}\label{scattering}
\mathrm{A}^{\mathrm{MHV}}_{n}\mathcal{A}_{n,\hat{k}}(\tilde{\lambda}_a,\tilde{\mu}_a,\tilde{\mathtt{z}}_{a\hat{\alpha}}).
\end{equation}
By substituting the corresponding relations \eqref{4-1cell-de} and
\eqref{4-diff-1cell-de} into expression \eqref{scattering}, it
gives a simple contour integration that reads
\begin{eqnarray}\label{scattering--}
&\!\!\sum\limits_{\Gamma\subset T}
         \frac{\mathrm{A}^{\mathrm{MHV}}_{n}
        <\tilde{Z}_{a_{(1)}},\cdots,
        \tilde{Z}_{a_{(4+\hat{k})}}\!>^4}
    {\!\!<\mathsf{Y}_0,\tilde{Z}_{a_{(1)}}\!,\tilde{Z}_{a_{(2)}}\!,\tilde{Z}_{a_{(3)}}
    \!,\tilde{Z}_{a_{(4)}}\!>\cdots<\mathsf{Y}_0,\tilde{Z}_{a_{(4+\hat{k})}}
  \!,\tilde{Z}_{a_{(1)}}\!,\tilde{Z}_{a_{(2)}}\!,\tilde{Z}_{a_{(3)}}\!>},\cr&
\end{eqnarray}
where
$\tilde{Z}_a\equiv(\vartheta_{a1},\cdots,\vartheta_{a4},\tilde{\mathtt{z}}_{a1},\cdots,\tilde{\mathtt{z}}_{a\hat{
k}})$). The simplest case of expression \eqref{scattering--} is
when $k=3$, which is equivalent to the one mentioned in
Ref.~\cite{Elvang-2015} except that
$\tilde{\mathtt{z}}_{a\hat{\alpha}} $'s have been replaced by
fermionic variables similar to relation~\eqref{bosonization}.

In
Refs.~\cite{Elvang-2015,Arkani-Hamed-2014,Arkani-Hamed-2016,Arkani-Hamed-2010},
it has been indicated that expression \eqref{scattering} is the
scattering amplitude of $n$ particles with total helicity $k$,
wherein $\mathrm{A}^{\mathrm{MHV}}_{n}$ is the maximally helicity
violating amplitudes (MHV) and
$\mathcal{A}_{n,\hat{k}}(\tilde{\lambda}_a,\tilde{\mu}_a,\tilde{\mathtt{z}}_{a\hat{\alpha}})$
is the amplituhedron in
$(\tilde{\lambda}_a,\tilde{\mu}_a,\tilde{\mathtt{z}}_{a\hat{\alpha}})$
space. Actually, through the resulted expression
\eqref{scattering}, we have introduced a sort of general structure
in comparison with the conventional amplituhedron, of which the
common amplituhedron presented in
Refs.~\cite{Elvang-2015,Arkani-Hamed-2014,Arkani-Hamed-2016,Arkani-Hamed-2010}
can be regarded as a particular choice~\eqref{bosonization}.
Indeed, in expression \eqref{scattering}, if the variables
$\tilde{\mathtt{z}}_{a\hat{\alpha}}$ are replaced by the
multiplication of two fermionic variables (named
bosonization~\cite{Arkani-Hamed-2018}), one will arrive at the
same common amplituhedron presented in
Refs.~\cite{Elvang-2015,Arkani-Hamed-2014,Arkani-Hamed-2016,Arkani-Hamed-2010}.

\section{Four-Dimensional Loop-Amplituhedron}

To get the loop-amplituhedron in a four-dimensional spacetime, we
employ the second case of definition \eqref{y} and, without loss
of generality, consider that $(k-l)$ of $y_\alpha$'s are equal to
zero. Also, we again assume that $z_a\in\mathbf{M}$. Thus, we get
a $(4+k-l)$-plane, wherein we choose $(4+k-l)$ independent
$n$-component column-vectors
$Z_a\equiv(\zeta_{a1},\cdots,\zeta_{a4},\mathtt{z}_{a1},\cdots,\mathtt{z}_{a,k-l})$
as a basis for it. Hence, expression \eqref{4-vol-hie} changes as
\begin{equation}\label{4-loop-vol-hie}
\int\dfrac{d^{k\times n}c_{\alpha
a}}{\prod\limits_{b=1}^nM_b}\delta^{[k\times(4+k-l)-4l]}(\mathtt{Y}-\mathtt{Y}_0)\equiv
\mathcal{A}_{n,k-l,l},
\end{equation}
where, here, the corresponding $\mathtt{Y}$ is a $ [k\times
(4+k-l)]$-matrix with constraint
\begin{equation}
\mathtt{Y}_0 = \begin{pmatrix}
       0_{(k-l)\times
4} & \mathbb{I}_{(k-l)\times (k-l)}  \\[0.3em]
       0_{l\times
4} & 0_{l\times(k-l)
}
     \end{pmatrix}.
\end{equation}

To proceed, due to relations \eqref{4-1cell-de} and
\eqref{4-diff-1cell-de} for the $C$ matrix, we need to increase
the number of components of the corresponding vectors $Z_a$'s from
$(4+k-l)$ to $(4+k)$. Accordingly, we define soft row-vectors
$Z'_{n+i}$'s, with $(4+k)$-component for $i=1,\cdots, l$, and new
$(4+k)$-component vectors $Z'_{a}$'s, for each $a$, through their
components as
\begin{equation}\label{4soft-vect}
Z'_{a}=(\overbrace{Z_{a}}^{4+k-l}
,\overbrace{0,\cdots,0}^{l})\quad {\rm and} \quad
Z'_{n+i,\gamma}=\delta_{4+k-l+i,\gamma},
\end{equation}
for $\gamma =1,\cdots,4+k$, which construct the corresponding
matrix $Z'$. Thus, we can write expression \eqref{4-loop-vol-hie}
as
\begin{equation}\label{4-new-loop-vol-hie}
\mathcal{A}_{n,k-l,l}=\int\dfrac{d^{k\times (n+l)}c_{\alpha
A}'}{\prod\limits_{B=1}^{n+l}M_B}\delta^{[k\times(4+k)-4l]}(\mathtt{Y'}-\mathtt{Y}_0'),
\end{equation}
where $A=1,\cdots,n+l$, arbitrary elements $c'_{\alpha A}$
constitute the corresponding Grassmannian $[k\times (n+l)]$-matrix
(say $C'$), $ M_B $'s are independent minors of $C'$,
$\mathtt{Y}'$ is a $[k\times (4+k)]$-matrix generally defined as $
\mathtt{Y}'\equiv C'\cdot Z'$ with constraint $\mathtt{Y}_0'
=(0_{k\times 4},\mathbb{I}_{k\times k})$. By comparing expressions
\eqref{vol-hie} and \eqref{4-new-loop-vol-hie}, it results that
$k'$ in \eqref{vol-hie} corresponds to $(k-l)$ in
\eqref{4-new-loop-vol-hie}.

In addition, using the approach presented in the previous section,
$\mathcal{A}_{n,k-l,l}$ can be transformed in the momentum twistor
space as
\begin{equation}\label{loop-scattering}
    \mathrm{A}^{\mathrm{MHV}}_{n}\mathcal{A}_{n,\hat{k}-l,l}
    (\tilde{\lambda}_a,\tilde{\mu}_a,\tilde{\mathtt{z}}_{a\hat{\alpha}}).
\end{equation}
With appropriate contour integration, expression
\eqref{loop-scattering} for a few fixed values of $n$, $k$ and
$l$, becomes
$\mathrm{A}^{\mathrm{MHV}}_{n}\mathcal{A}_{n,\hat{k}-l}$.
Moreover, for $l$ as even numbers, four-dimensional
loop-amplituhedron arises from the `entangled'
integration~\cite{Arkani-Hamed-2011} of each pair of
$(\mathtt{Y}_{k-l+1},\cdots,\mathtt{Y}_k)$ with $l/2$ loops.

As an example, for a simple contour integration of
\eqref{4-new-loop-vol-hie}, using relations
\eqref{4-1cell-de} and
\eqref{4-diff-1cell-de}, with Grassmannian\footref{ffoott}
\begin{widetext}
\begin{equation}\label{matrix4l}
    \begin{pmatrix}
        c_{1,a_{(1)}} & \cdots & c_{1,a_{(4+k-l)}} & c_{1,a_{(5+k-l)}} & \cdots & c_{1,a_{(4+k)}}\\
        \vdots & \ddots & \vdots & \vdots & \ddots & \vdots \\
        c_{k-l,a_{(1)}}& \cdots & c_{k-l,a_{(4+k-l)}} & c_{k-l,a_{(5+k-l)}}& \cdots & c_{k-l,a_{(4+k)}} \\
            c_{k+1-l,a_{(1)}} & \cdots & c_{k+1-l,a_{(4+k-l)}} & c_{k+1-l,a_{(5+k-l)}} & \cdots & c_{k+1-l,a_{(4+k)}} \\
        \vdots & \ddots & \vdots & \vdots & \ddots & \vdots \\
        c_{k,a_{(1)}}& \cdots & c_{k,a_{(4+k-l)}} & c_{k,a_{(5+k-l)}}& \cdots & c_{k,a_{(4+k)}}
    \end{pmatrix},
\end{equation}
\end{widetext}
into expression \eqref{4-new-loop-vol-hie}, it reads
\begin{eqnarray}\label{4-loop-hier-pha-spa}
&\!\!\sum\limits_{\Gamma\subset T}\!\int \!\!\frac{
d^{k\times(4+k)}\mathtt{Y}_\alpha'\!<Z_{A_{(1)}}',\cdots,
Z_{A_{(4+k)}}'\!>^4\delta^{[k\times(4+k)-4l]}(\mathtt{Y}'\!-\mathtt{Y}_0')}
{\!<\mathsf{Y}'\!,Z_{A_{(1)}}'\!,Z_{A_{(2)}}'\!,Z_{A_{(3)}}'\!,Z_{A_{(4)}}'\!\!\!>\cdots<\mathsf{Y}'
\!,Z_{A_{(4+k)}}'\!,Z_{A_{(1)}}'\!,Z_{A_{(2)}}'\!,Z_{A_{(3)}}'\!\!>},\cr&
\end{eqnarray}
where each one of $A_{(i)}$ gets a different number from
$(1,\cdots,n+l)$,
$\mathsf{Y}'\equiv\mathtt{Y}_1'\wedge\cdots\wedge\mathtt{Y}_k'$
and each $\mathtt{Y}_\alpha' $ is $\alpha$th row-vector of the
matrix $\mathtt{Y}'$ with $(4+k)$ components. By substituting
components \eqref{4soft-vect} into expression
\eqref{4-loop-hier-pha-spa} and integrating over the components of
the corresponding matrix $\mathtt{y}'$ (namely,
$\mathtt{y}_{\alpha\beta}'$ for $\alpha =k-l+1,\cdots,k$ and
$\beta =1,\cdots,k$), and after some manipulations, it becomes
\begin{equation}\label{mod-4-loop-hier-pha-spa}
\scriptsize{\!\!\!\!\!\!\sum\limits_{\Gamma\subset
T}\!\int\!\frac{
d^{k\times(4+k-l)}\mathtt{Y}_\alpha\!<Z_{a_{(1)}},\cdots,
Z_{a_{(4+k-l)}}\!>^4\delta^{[k\times(4+k-l)-4l]}(\mathtt{Y}\!-\!\mathtt{Y}_0)}
{\!\!\!<\mathsf{Y}\!,Z_{a_{(1)}}\!,Z_{a_{(2)}}\!,Z_{a_{(3)}}\!,Z_{a_{(4)}}\!>\cdots<\mathsf{Y},
\mathtt{Y}_k,Z_{a_{(1)}}\!,Z_{a_{(2)}}\!,Z_{a_{(3)}}\!>},}
\end{equation}
where in here
$\mathsf{Y}\equiv\mathtt{Y}_1\wedge\cdots\wedge\mathtt{Y}_{k-l}$.

For instance, the simplest case of expression
\eqref{mod-4-loop-hier-pha-spa} in the momentum twistor space is
when $n=4$, $l=2$ and $k=4$ (and, in turn, $\hat{k}=2$). Also, by
employing the corresponding expression
\eqref{mod-4-loop-hier-pha-spa} for $\tilde{Z}_a$ and $\hat{k}$
instead of $Z_a$ and $k$, and integrating over the corresponding
delta functions, the result reads
\begin{widetext}
\begin{equation}\label{exam-mod-4-loop-hier-pha-spa}
    \mathrm{A}^{\mathrm{MHV}}_{4} \int \frac{
        d^4y_1d^4y_2<\tilde{Z}_1,\cdots,
        \tilde{Z}_4>^3}
{<\tilde{Z}_2,\tilde{Z}_3,\tilde{Z}_4,\mathtt{Y}_{1}><\!\tilde{Z}_3,\tilde{Z}_4,\mathtt{Y}_{1},\mathtt{Y}_{2}><\tilde{Z}_4,
\mathtt{Y}_{1},\mathtt{Y}_{2},\tilde{Z}_1><\mathtt{Y}_{1},\mathtt{Y}_{2},
\tilde{Z}_1,\tilde{Z}_2><\mathtt{Y}_{2},\tilde{Z}_1,\tilde{Z}_2,\tilde{Z}_3>}.
\end{equation}
\end{widetext}
Then, by using the `entangled' contour of
integration~\cite{Arkani-Hamed-2011}, result of
\eqref{exam-mod-4-loop-hier-pha-spa} gives
\begin{widetext}
\begin{equation}\label{exam1-mod-4-loop-hier-pha-spa}
    \mathrm{A}^{\mathrm{MHV}}_{4} \int \left[ \frac{d^4y_1d^4y_2}{\mathrm{vol}[GL(2)]}\right] \frac{
        <\tilde{Z}_1,\cdots,
        \tilde{Z}_4>^2}
{<\mathtt{Y}_{1},\mathtt{Y}_{2},\tilde{Z}_1,\tilde{Z}_2><\mathtt{Y}_{1},
\mathtt{Y}_{2},\tilde{Z}_2,\tilde{Z}_3><\mathtt{Y}_{1},\mathtt{Y}_{2},
\tilde{Z}_3,\tilde{Z}_4><\mathtt{Y}_{1},\mathtt{Y}_{2},\tilde{Z}_4,\tilde{Z}_1>},
\end{equation}
\end{widetext}
that is equivalent to the one mentioned in
Refs.~\cite{Elvang-2015,Arkani-Hamed-2011} except that the
variables $\tilde{\mathtt{z}}_{a\hat{\alpha}} $'s must be replaced
with the fermionic variables. Actually, as in the previous
section, through the resulted expression
\eqref{exam1-mod-4-loop-hier-pha-spa}, a sort of general structure
in comparison with the conventional loop-amplituhedron has been
introduced, of which the common amplituhedron presented in
Refs.~\cite{Elvang-2015,Arkani-Hamed-2011} can be regarded as a
particular choice~\eqref{bosonization}.

\section{Discussion}

\textbf{Physical Description of Closed Contour Link Mode}:$\,{} $
In order to attain some view on the physical description of the
closed contour link mode, let us give some examples. Consider
schematically the Feynman diagram of the interaction of two
particles in a massless theory as
\begin{figure}[htbp]
    \includegraphics[width=8cm]{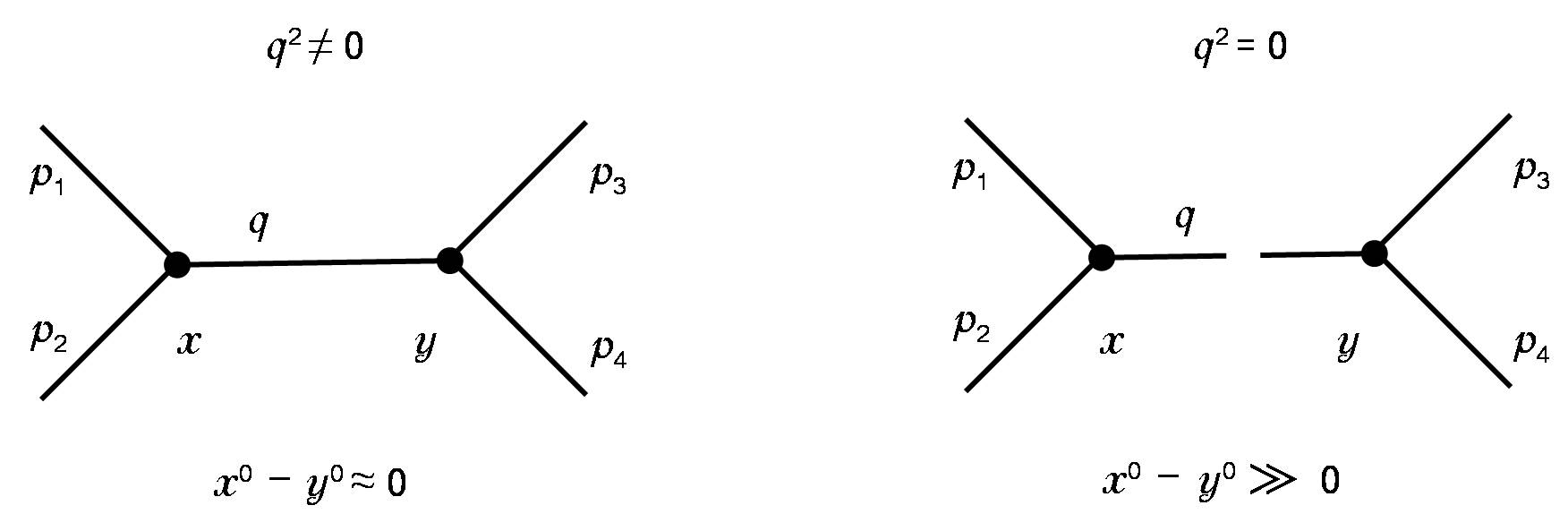}
    \caption{\label{fig2}\small The schematic Feynman diagrams of the interaction of
    two particles in a massless theory.}
\end{figure}
shown in Fig.~\ref{fig2}. When the time interval between vertexes
$x$ and $y$ is very short, the magnitude of the energy-momentum
vector for the propagator is non-zero and hence, the propagator is
a virtual particle. However, when the time interval between these
points is very long, the magnitude of the energy-momentum vector
for the propagator is zero and hence, the propagator is a real
particle. Therefore, for the first case, one needs to have one
closed contour for real particles, while for the second case, two
closed contours for real particles are required. Now consider
another example in which five groups of particles interact with
each other in a massless theory. If these five groups of particles
move away from each other in such a way that the distances between
those increase very much, then the propagators among those will
become real particles (see, e.g., Fig.~\ref{fig3}).
\begin{figure}[htbp]
 \includegraphics[width=7.5cm]{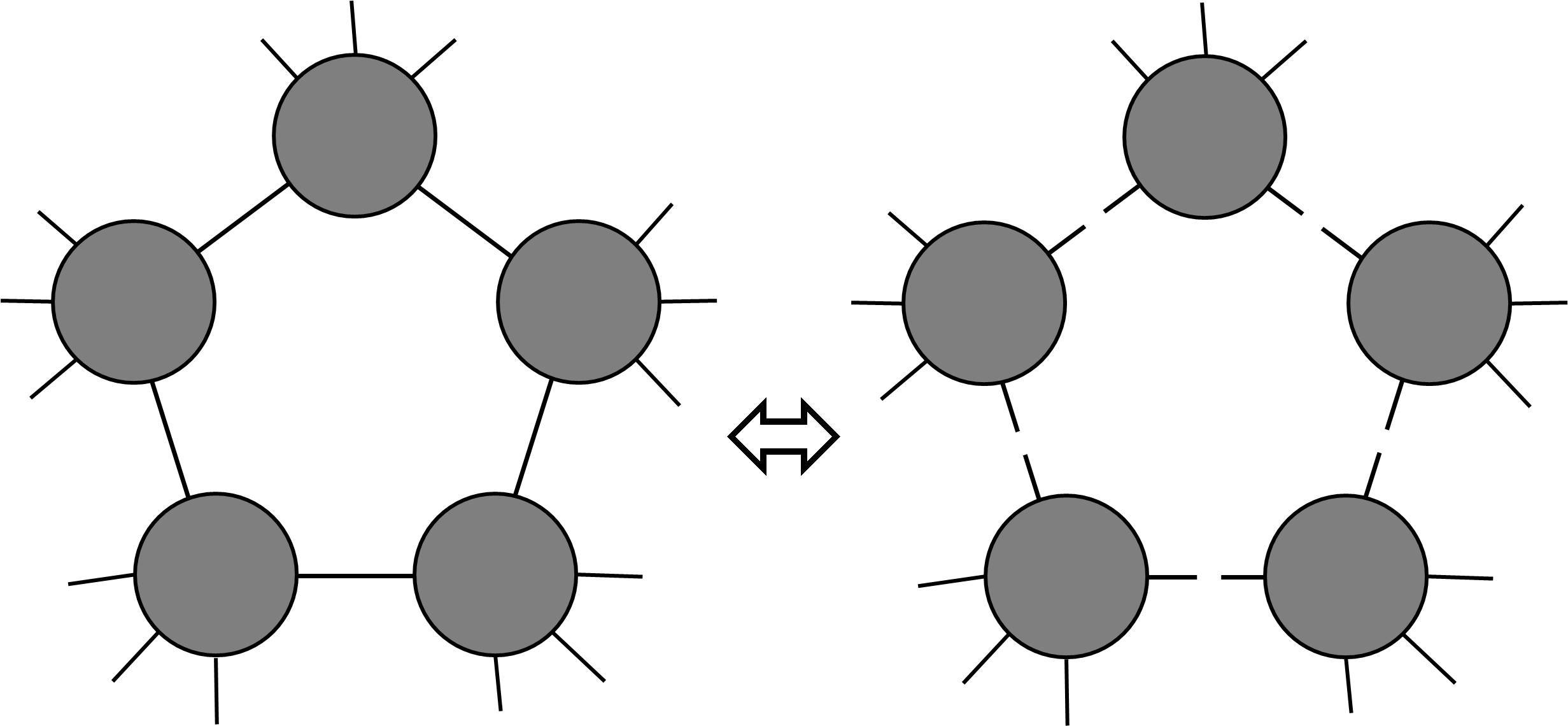}
 \caption{\label{fig3}\small The schematic Feynman diagrams of the interaction of five groups of particles in a massless theory.}
\end{figure}
Each group of particles behaves like a composite particle and also
interacts with other groups in a way that the outcome forms a
closed contour of the momentum conservation. For example, see
Fig.~\ref{fig4}, wherein
\begin{figure}[htbp]
    \includegraphics[width=7.5cm]{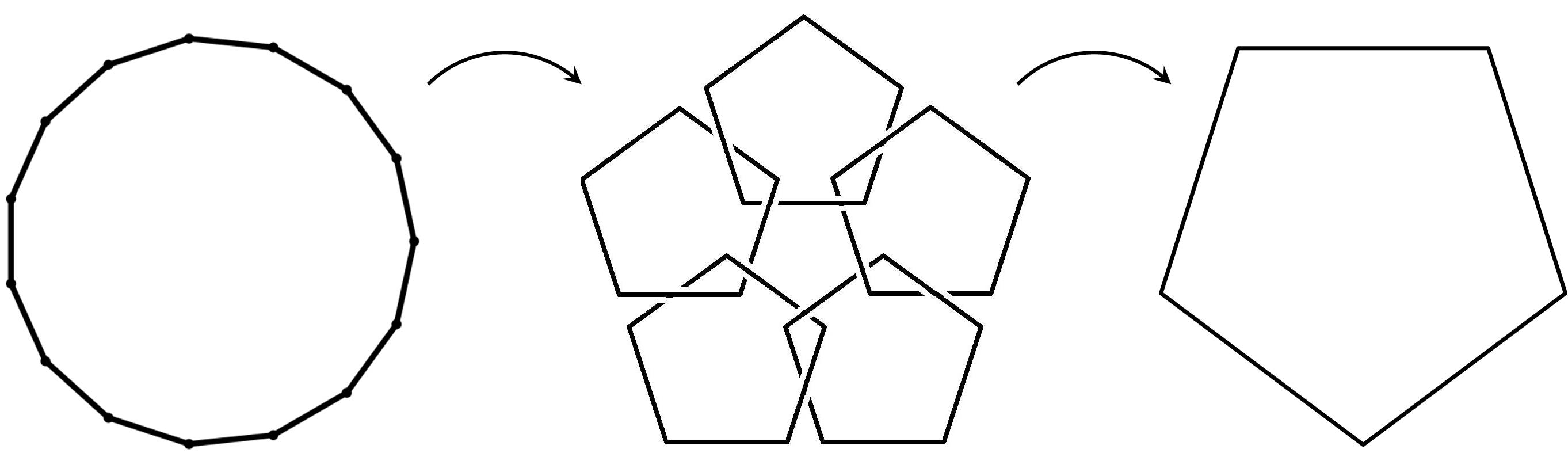}
    \caption{\label{fig4}\small The closed contour of the momentum conservation of five
    groups of particles that move away from each other in such a way that the
    propagators among those become real particles, related to Fig.~\ref{fig3}.}
\end{figure}
the reason for linking these five closed contours in this figure
is that these five sets of particles are~not independent of each
other and interact with each other so that the result forms a
closed contour of the momentum conservation. In the standard field
theory, there is no~significant physical difference between the
left and right parts of Fig.~\ref{fig3}. However, in the PT,
geometrical difference of the left and right parts of
Fig.~\ref{fig3} leads to different topology, and hence, physical
properties.

Since links and knots have topological properties, we expect a
correspondence between these topological properties, and in turn,
the properties of the consisted structure of particles and
interactions among those. For example, the carbon nanotubes,
graphite and diamond are all made of carbon but have completely
different properties. It is widely accepted that the difference in
their properties is due to their different geometric structures in
the placements of atoms of carbon. As an another example, let us
mention the topological molecules~\cite{Diudea}, in which those
atoms, that make up some of the molecules, may be the same but
have completely different topological
structures~\cite{Forgan,Shimokawa}.
 \vskip0.5cm

\textbf{PT Description of Loop Level Scattering Amplitude}:$\,{} $
From mathematical point of view, the result of the hidden
particles method in describing loop corrections (mentioned in
Refs.~\cite{Elvang-2015,Arkani-Hamed-2011}) is equivalent to what
we have mentioned in the previous section. In fact, in the hidden
particles method, two hidden particles are considered for each
loop, and their momentum twisters are integrated in a certain way.
On the other hand, in the previous section, instead of integrating
on the momentum twisters of two hidden particles, the same number
of Dirac delta functions have been removed and integrated on
variables $y_1$ and $y_2$ in the same specific way.

\begin{figure}[htbp]
    \includegraphics[width=7.5cm]{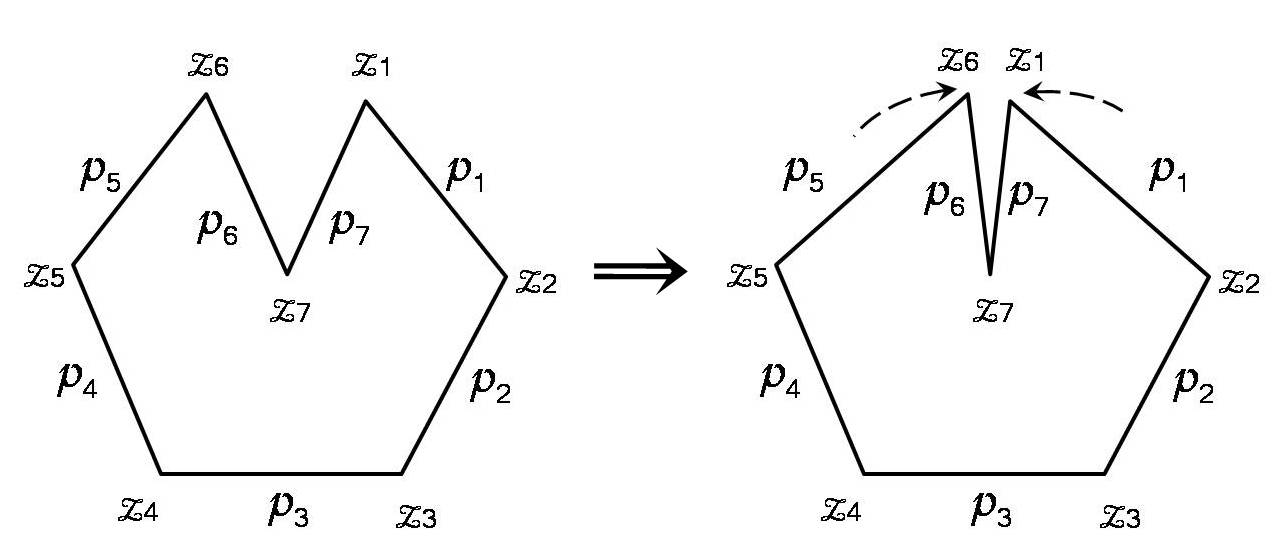}
    \caption{\label{fig5}\small Forming a loop in the closed contour of the momentum conservation.
    After the points $z_1$ and $z_6$ touch each other, the energy-momentum vectors $p_6$ and $p_7$
    will be equal and in opposite directions~\cite{Elvang-2015}.}
\end{figure}

Physically, from the PT point of view (also from the point of view
of on-shell physics~\cite{Elvang-2015,Arkani-Hamed-2011}), a loop
means the existence of a pair of entangled real
particle-antiparticle whose energy-momentum vectors are opposite
to each other. Thus, the existence of these pairs does~not change
the total energy-momentum, and therefore all possible values for
their energy-momentum must be considered. This state is
schematically as a state in which two points of the
one-dimensional manifold touch each other~\cite{Elvang-2015} (see,
e.g., Fig.~\ref{fig5}).

\section{Conclusions}

Inspired by the fact that geometric representation of the momentum
conservation of particles participating in an interaction leads to
the formation of a closed contour, we have introduced an
integrable one-dimensional theory, i.e., the PT. This theory
underlies some integrable models such as the KP-hierarchy and the
amplituhedron. In this regard, we have defined the action and then
the partition function of the PT. Afterward, we have shown that
the fields of the PT are equivalent with their partition
functions. Due to the quantization process, this issue means that
the theory will~not change after quantization. Therefore, this
theory possesses a particular symmetry as a statistical symmetry.
This symmetry implies that the theory, in small-scales, is
statistically equivalent to large-scales (such a characteristic is
what we have benefited in another work~\cite{you-far}).

Also, by defining the S-matrix of the PT, we have indicated that
the S-matrix has the Grassmannian structure. This Grassmannian
corresponds to a chord diagram, which specifies the closed contour
of the PT. Thereupon, by introducing and using a transformation
(i.e., the hierarchy transformation), we have shown that the PT
underlies some important kinds of integrable models in different
dimensions such as the types of soliton models derived from the
KP-hierarchy, two-dimensional models and the amplituhedron. Thus,
we have represented that the S-matrix of the PT, after hierarchy
transformation, is a solution of the KP-hierarchy.

Furthermore, we have indicated that the volume of phase-space of
the one-dimensional manifold of the theory is equivalent with the
Grassmannian integral. Indeed, contrary to
Refs.~\cite{Elvang-2015,Arkani-Hamed-2014,Arkani-Hamed-2016,Arkani-Hamed-2011},
without any use of supersymmetry, we have obtained a sort of
general structure in comparison with the conventional Grassmannian
integral and the resulted amplituhedron, which is closely related
to the Yang-Mills scattering amplitudes in four dimensions.
Accordingly, the common amplituhedron presented in the literature
can be derived from the volume of phase-space of the proposed
one-dimensional manifold as a special contour integration of the
Grassmannian integral presented in this work, and being regarded
as a particular choice of it. Indeed, the proposed theory is
capable to express both the tree- and loop-levels amplituhedron
(without employing hidden particles) in four dimensions in the
twistor space. Finally, we have discussed the physical meaning and
importance of the linked closed contour.

\section*{Acknowledgements}

We thank the Research Council of Shahid Beheshti University.
%%%%%%%%%%%%%%%%%%%%%%%%%%%%%%%%%% References %%%%%%%%%%%%%%%%%%%%%%%%%%%%%%%%%%

%
\end{document}